\documentclass{article}

\usepackage{arxiv}
\usepackage{orcidlink}

\usepackage[utf8]{inputenc} % allow utf-8 input
\usepackage[T1]{fontenc}    % use 8-bit T1 fonts
\usepackage{hyperref}       % hyperlinks
\usepackage{url}            % simple URL typesetting
\usepackage{booktabs}       % professional-quality tables
\usepackage{amsfonts}       % blackboard math symbols
\usepackage{nicefrac}       % compact symbols for 1/2, etc.
\usepackage{microtype}      % microtypography
\usepackage{lipsum}
\usepackage{graphicx}
\graphicspath{ {./images/} }

\title{Ownership and Flow Primitives for Scalable Consent Management in Digital Public Infrastructures}

\author{
Rohith Vaidyanathan\textsuperscript{\orcidlink{0009-0000-7021-2787}} \\
Web Science Lab, IIIT Bangalore\\
Bengaluru, Karnataka, India \\
\texttt{rohith.vaidyanathan@iiitb.ac.in} \\
\And
Srinath Srinivasa\textsuperscript{\orcidlink{0000-0001-9588-6550}} \\
Web Science Lab, IIIT Bangalore\\
Bengaluru, Karnataka, India \\
\texttt{sri@iiitb.ac.in} \\
\And
Praseeda\textsuperscript{\orcidlink{0009-0008-8244-5084}} \\
Web Science Lab, IIIT Bangalore\\
Bengaluru, Karnataka, India \\
\texttt{praseeda@iiitb.ac.in} \\
\And
Dev Shinde\textsuperscript{\orcidlink{0009-0007-3016-2181}} \\
Web Science Lab, IIIT Bangalore\\
Bengaluru, Karnataka, India \\
\texttt{shindedev.hemravi@iiitb.ac.in}
}
\begin{document}
\maketitle

\begin{abstract}
  Digital public infrastructures (DPIs) represent networks of open technology standards, applications, services, and digital assets made available for the public good. One of the key challenges in DPI design is to resolve complex issues of consent, scaled over large populations. While the primary objective of consent management is to empower the data owner, ownership itself can come with variegated morphological forms with different implications over consent. Questions of ownership in a public space also have several nuances where individual autonomy needs to be balanced with public well-being and national sovereignty. This requires consent management to be compliant with applicable regulations for data sharing. This paper addresses the question of representing modes of ownership of digital assets and their corresponding implications for consensual data flows in a DPI. It proposes a set of foundational abstractions to represent them. Our proposed architecture responds to the growing need for transparent, secure, and user-centric consent management within Digital Public Infrastructure (DPI). Incorporating a formalised data ownership model enables end-to-end traceability of consent, fine-grained control over data sharing, and alignment with evolving legal and regulatory frameworks. \end{abstract}

% keywords can be removed
\keywords{Digital Public Infrastructure, Consent Management, Data Ownership, Data Flow Control, Computers and Society, Privacy Architecture, Data Governance
}

\section{Introduction}

Digital Public Infrastructures (DPIs)~\cite{undp_dpi} represent open ecosystems of digital services, applications and assets made available for public good. DPI services like digital identity, lockers, catalogs, wallets and payment infrastructures have not only streamlined several e-governance activities, but also a variety of economic, legal and social transactions among ordinary citizens~\cite{bandura2023unpacking}. There have been several initiatives in the past where, digital assets and services were designed to be owned by and available for the public like free software movements, creative commons, etc. However, the term DPI refers to systemic, scalable \textit{infrastructure} that is meant to act as a ``digital backbone'' for the entire society, enabling access to both public and private services like healthcare, education, and financial inclusion. The role of the state and other institutional players is central to the design of DPIs, in contrast to a predominantly community-driven approach in the earlier examples. 

As public interactions become more digital, vast amounts of sensitive data, including personal data, are exchanged through DPIs as a necessary step to provide services. But without proper mechanisms in place, data owners may lose control and knowledge of how their sensitive data is used post exchange. In response to these challenges, various countries have implemented data protection and consent management laws to protect individuals and institutions~\cite{gabriela2018eu}. Consent management is modeled as an interception of a data flow in a DPI, to ensure that the flow is consistent with the consent of the data owner.

% DPI promotes interoperability among various systems and sectors, allowing for seamless data exchange and collaboration between public and private entities, enabling comprehensive solutions for complex societal challenges~\cite{bandura2023unpacking}.

% (DEPA)\footnote{\url{https://www.niti.gov.in/sites/default/files/2020-09/DEPA-Book.pdf}}.  

% \begin{figure}
% \centering
% \includegraphics[width=3in]{depa.png}
% \caption{DEPA consent architecture from \url{https://www.niti.gov.in/sites/default/files/2020-09/DEPA-Book.pdf}}
% \label{fig:depa} 
% \end{figure}

In its simplest form, a consent architecture elicits explicit consent from the Data Owner (DO) in response to a data request. This is called the Autonomous Authorization (AA) model of consent management~\cite{schermer2014crisis}. The AA model is suitable for cases where the data owner is directly involved in regulating the flow of their data, and the consent requests are not too frequent. But, as DPI implementations scale across populations, the AA architecture would be insufficient. Too frequent requests may lead to consent desensitization, where the data owner may not adequately understand the implications of her choices. Similarly, if data is shared by an independent custodian on behalf of the owner, the custodian cannot autonomously approve or deny requests-- and will need to adhere to some kind of data sharing policy set by the owner. The same is true, if the data owner is an institution, rather than an individual. 

Consent management also has several other nuances. Consent is closely related to ownership. One can give consent only on what one owns. But ownership itself can have several morphological forms. Ownership may be \textit{delegated} or \textit{pledged}, and partial ownership or privileges can be \textit{conferred}-- all of which, have implications on how consent may be granted and enforced. Similarly, consent management includes, but goes beyond, issues of access control. While there are several architectures for flexible and scalable access control, consent provisions may need to be enforced even after access is granted. Some examples of such post-conditions include consenting to whether the data may be shared further, aggregated with other data, the purposes for which they may be used, and requirements of suitable notifications to the data owner for any downstream use of their data. 

Finally, concerns about public well-being and national security and sovereignty would override any policies to the contrary, issued by the data owner. For instance, public health regulations may require certain contagious conditions to be reported by people to certain authorities, even if it goes against their individual sense of privacy. 

There is a need for a scalable architectural solution that would enable consensual data flows in DPIs by resolving issues of ownership, access policies, regulatory compliance, and assertion of data ownership post sharing. In this paper, we address this issue, and propose an architecture for DPIs that address the complex forms of ownership and flow primitives required for supporting consensual data flows.

\section{Related Literature}
\label{sec:related}

Although the term DPI is a fairly recent introduction, some of its core data-related challenges like inter-organizational, and open-ended sharing of sensitive data, have been addressed since several years~\cite{houser2022data,shrivastava2023comprehensive,stalla2020data,tith2020patient}. Inter-organizational data sharing addresses workflows that span across multiple organizations and governance structures, and the potential risks involved~\cite{van2015modes,JUSSEN2024102280}. Conventionally, access-control is enforced within organizational boundaries, using a consistent system of logic to resolve access requests. When access requests need to cross organizational boundaries and their corresponding governance structures, it could potentially lead to unsafe data access. Innovations in this area include different extensions of Role-Based Access Control (RBAC) models~\cite{park2001role,abdunabi2013specification,srinivasa2020legitimate}, Attribute-based models~\cite{hu2013guide,hu2015attribute} or hybrid models~\cite{bs2021towards,hybridRBAC2022}.

Issues of consent management in public data flows, go beyond just access control~\cite{balambiga2024cods}. The question of consent is closely related to issues of ownership, and the consideration of individual autonomy and  collective well-being. Research efforts related to consent management have addressed philosophical issues of what makes consent meaningful~\cite{williams2017consent,saksena2021rebooting,tschider2020meaningful,schermer2014crisis,karandikar2024makes}, as well as various consent architectures for different kinds of data sharing environments~\cite{balambiga2024cods,ulbricht2016,mont2012,pearson2011sticky}. 

%Existing consent management systems primarily treat consent from an access control lens. Blockchain-based approaches like~\cite{MKK24} focus on grant/revoke operations through smart contracts with immutable blockchain records, while graph-based systems like~\cite{10.1145/3665252.3665265} used to optimize data flows by disconnecting vertices Eg "User doesn't want location data used for ads → disconnect location vertex from advertising vertex". However, these approaches view consent primarily as a gateway issue, neglecting critical aspects like data ownership, user awareness of post-sharing data use, and autonomy to track and revoke access. Effective consent management should empower users with continuous control beyond the initial consent decision.

Existing consent management systems such as DICON~\cite{9881506} and the CMA for Secure Data Exchange framework~\cite{inproceedings2} emphasize policy and protocol based consent enforcement. DICON implements consent as a compliance mechanism using semantic ontologies and rule validation, while CMA adopts a MyData-inspired, operator-mediated workflow to authorize data transfers. In both models, consent primarily functions as a conventional authorization checkpoint rather than as a dynamic or user-centric control.

While frameworks like W3C Privacy Principles~\cite{W3C} provide valuable guidelines for web-based consent interactions, there is still a need for architectural solutions to design and implement consensual data flows in DPIs, that address complex ownership morphologies and institutional governance. Most existing consent managers employ \textit{Autonomous Authorization (AA)}~\cite{schermer2014crisis}, which is not scalable. The Fair Transaction Model, proposes a concept of \textit{legitimately implied consent}, to identify cases of consent without explicit authorization~\cite{miller2010preface,miller2011fair}. Policy-based consent~\cite{balambiga2024cods}, which is an implementation of the fair transaction model, evaluates and enforces regulations and data sharing policies in a domain agnostic manner for legitimate data flow. Certain conditions of consent may need to be enforced even after data has been shared. Some examples of such post-conditions include restrictions on whether the data can be downloaded or re-shared by the requester, how many times the data may be used, whether the data can be aggregated with other datasets, etc. There are also deeper philosophical questions pertaining to ownership that data sharing raises. Researchers~\cite{asswad2021,Hart_2002} have addressed questions like: Does sharing of data constitute sharing or transferring of ownership to the recipient, and what data flows can one legitimately control by virtue of ownership. Given data flows care non-rivalrous in nature (that is, data flows are non-conserved-- sharing a piece of data by a sender will not result in the sender losing that data from its end), these questions challenge conventional ideas about possession and highlight the need to reconsider what ownership and control mean in the digital age.

The idea of DPI makes consent management domain and organization agnostic, and makes the role of the regulator much more central. Regulations in India~\cite{DPDPA2023} and Europe~\cite{He2023datamarketplaces} have outlined mechanisms of consent to process open-ended public data lawfully. One of the pioneering frameworks addressing consent management in public data flows, is X-road~\cite{xroad_technology_overview,kalja2002x} which was started in Estonia, and other similar efforts include~\cite{yli2018suomi,hay2014national}. Some DPIs have introduced data ownership registration systems to certify the different rights of the data owner~\cite{He2023datamarketplaces}. Ownership over digital assets is also addressed in the Web3 paradigm with the concept of NFTs (Non-fungible tokens). However, while an NFT represents a title deed over a digital asset much like in the real world, it does not by itself confer any specific legal rights or entitlements~\cite{murray2022nft}. 

%Storing an NFT on a permissioned blockchain can be an enabling mechanism for asserting ownership, but there is still a need for representing and enforcing the semantics and legal implications of ownership and its morphological forms (delegated, leased ownership etc) as digital assets are exchanged in an open-ended fashion on DPIs. In order to address these, we need an extended notion of data ownership itself. Currently, most definitions of data ownership assume that the data owner has unfettered access over datasets owned by them. 

As data is shared in an open-ended fashion in DPIs, we argue that data ownership morphs into different constructs, which need to be meaningfully modeled regarding their impact on consent. An effective consent management system must encompass not only access control, but also explicitly address issues of ownership to enable ongoing user control over data following its sharing, and ensure compliance with evolving data policies. In this paper, we propose a comprehensive model that systematically addresses all four of these critical dimensions of consent management.

% \begin{table}[t]
% \caption{\label{tab:definitions} 
% Essential DPI definitions}
% \begin{tabular}{|p{4cm}|p{10cm}|}
% \hline
% Term & Definition\\
% \hline
% \hline 
% Data Owner (DO) & Owner of a given dataset\\ 
% Data Principal (DP) & Entity authorized to give consent on a given dataset (may or may not be the same as the data owner)\\
% Data Custodian (DC) & Entity authorized to collect data from data owners and share on their behalf. May also play the role of a Data Principal\\ 
% Data Subject (DS) & Entity about which, the data is about\\
% Data Trust (DT) & A fiduciary entity that stores and shares data on behalf of its owner\\
% Data Requester (DR) & Entity requesting for a given dataset\\
% Data Processor (DPr) & Entity that processes a given dataset (may or may not be the same as the requester)\\
% \hline
% \end{tabular}
% \end{table}

\section{Modeling Public Data Flows} 
\label{sec:publicdataflow}

% In order to model public data flows, we need to define some essential roles involved in this process.

% The Data Owner (DO) is the entity that owns a data element. A DO can be either an individual or an institution. Sometimes, several data owners may give their data to be kept in confidence by a Data Custodian (DC) and also authorize the custodian to share the data with certain recipients on their behalf. The Data Principal (DP) is the entity that is the source of consent. Typically, the DP is the same as the DO, but in some cases, the DP could be the custodian who is acting on behalf of the DO. The Data Subject (DS) is the entity described by the data element. For personal data, typically the DS is the same as the DO. However, there can be cases where the subject is different from the owner of the data. For instance, when a university creates academic transcripts about a student, the university is the DO, while the student is the DS. Similarly, photographs of celebrities taken by a photographer are owned by the photographer, even though the subject is the celebrity. Finally, the entity requesting for a data element is the Data Requester (DR), who may in turn give it to a Data Processor (DPr) which may or may not be the same as the DR, for processing the data. Table~\ref{tab:definitions} provides a list of definitions the key roles that participate in a DPI, that are considered in this work. 

Any public infrastructure poses a complex interplay between entitlements of individuals and the public at large. While individuals and the public form two ends of a spectrum, in between, there can be communities, institutions and other organizational entities that stake their own ownership. 

Given this, we model a public data flow network to be made up of a number of semantic containers, each representing ownership boundaries of its stakeholders. Formally, data flows in a DPI can be modeled with the following components:

% Figure~\ref{fig:dpi-graph} schematically depicts this.   

% \begin{figure}
%     \centering
%     \includegraphics[width=3in]{multiverse.png}
%     \caption{Modeling Public Data Flows}
%     \label{fig:dpi-graph}
% \end{figure}

\begin{equation}
    DPI = (A, W, F) 
\end{equation}

Here, $A$ is a set of \textit{agents} or \textit{stakeholders} who assert ownership, and play different roles like a Data Owner (DO), Data Requester (DR) or a Data Subject (DS). $W$ is a set of containers representing semantic boundaries called as \textit{Access Policy Domain (APD)}, or more colloquially as \textit{world} or \textit{locker}. They represent semantic containers where data ownership is enforced. An agent may own one or more lockers each with its own set of policies. The term $F \subseteq W \times W$ represent data flow pipelines, also called \textit{Connections}, established between lockers that represent legitimate pathways by which, data can be exchanged between different stakeholders. Connections are made legitimate by an underlying contract encapsulating the data sharing policies of the respective stakeholders, as well as applicable regulations.

Lockers publish one or more \textit{connection endpoints} that are based on a contractual template. Other stakeholders visiting this locker can connect any one of their lockers to any applicable connection endpoint. The locker publishing a connection endpoint is called the \textit{host} of the connection, and a locker connecting to this endpoint is called the \textit{guest} of this connection. 

An artifact is a logical unit of data that is subject to ownership and consent and that flows through connections. It represents a data resource (e.g., a document or image), though a single resource may have multiple artifacts. The consent service (comprising agents, lockers, and connections) regulates artifact storage and flow, while a separate resource service manages actual resources. When a DR requests an artifact, lockers and connections resolve consent-related issues before requesting the resource service to provide access.

% A unit element that is subject to ownership and consent, and that flows on the connections is called an \textit{artifact}. An artifact is distinguished from a \textit{resource}, which is the actual data element (like a document or an image). Typically, an artifact represents a resource in its entirety, however, a resource may have several artifacts representing different parts of it. The entire consent service made up of agents, lockers, and connections is meant to regulate the storage and flow of artifacts in a legal and consensual manner. The actual resources are managed by a separate resource service. When a DR makes a request for an artifact, the artifact flow is first resolved using lockers and connections, following which, a request is made from a locker to the resource service to make the resource available to the DR. This process is elaborated in the next section. 

For any given locker $w \in W$ and agent $a \in A$, let $a$ be the owner of locker $w$, denoted as $owner(w,a)$. The agent $a$ may make available some artifacts in $w$, and also have $w$ connect to other lockers for data exchange. If agent $a$ wishes to access some artifacts that are available in locker $w'$ belonging to agent $a'$, then $a$ first connects to a suitable endpoint published by $w'$ presenting $w$ as the participating locker from its side. Once the agreement is established, this leads to the formation of links $(w,w'), (w',w) \in F$. All data flow on these links will need to adhere to the terms of the agreement that was mentioned when this connection was created. All data access happen only between a given agent and any locker that it owns. An artifact present in $w$ could represent a resource belonging to $a$ or something that has been acquired consensually from other lockers. Any application program built on top this DPI architecture will be treated as an agent, and will have its own set of lockers. 

The above system of lockers and connections regulate the flow of \textit{artifacts}, and not \textit{resources}. Artifacts are logical representations of resources (or parts of them) that enable the holder of the artifact to access the resource using the resource service. 

With this setup, we ask the following questions (i) How does the owner of a dataset establish its ownership on data elements that it has shared and are now residing in other lockers? (ii) How does the owner of a dataset enforce post conditions of data access, that address questions like for what purpose this data may be used, and what should be done with the data after its use? (iii) How does an agent reliably delegate responsibility to other agents to manage its data on its behalf?

\section{Consent Flow Architecture}
\label{sec:flow}

In this section, we illustrate the consent flow using a DPI using a running example. A student $s$ uses a DPI to get her degree from university $u$ and is now applying for a job with company $c$ using the same DPI. The example takes cue from the questions motivated above and illustrates multiple interactions: 

\begin{description}
\item[Degree granting:] Student $s$ requests a degree certificate and transcripts from university $u$. While $s$ is the owner of her degree and academic transcripts, she cannot modify them unilaterally, making her the \textit{conferred owner}, while the university $u$ that grants the degree, remains the \textit{primary owner}.

% Here, student $s$ approaches university $u$ to obtain her degree certificate and academic transcripts. The university that has created these documents is the Data Owner (DO) and the student is the Data Requester (DR). Since the certificate and the transcripts are about student $s$, she becomes the owner of these documents after they are issued. However, despite being the owner of her transcripts, student $s$ may not make unilateral changes to these documents. Student $s$ is hence, a \textit{conferred} owner of these documents, while university $u$ remains the \textit{primary} owner. 
\item[Job application:] Student $s$ shares her academic credentials with company $c$ for a job application. In such a case, the company only requires \textit{access rights} to these document rather than owning these documents after they are shared. While in the previous case ``sharing'' meant conferring ownership, in this case, ``sharing'' simply means granting access rights. 

% Student $s$ approaches company $c$ to apply for a job by presenting her certificates and transcripts. For the job application, the student will need to present her credentials. In such a case, the company only requires \textit{access rights} to these document rather than owning these documents after they are shared. While in the previous case ``sharing'' meant conferring ownership, in this case, ``sharing'' simply means granting access rights. 
\item[Credentials Verification:] Company $c$ seeks verification from university $u$, about the credentials of student $s$. The university may verify the credentials, without the consent of $s$ (since it is the primary owner) unless regulations require otherwise. Here too, data sharing only provides access privilege without any transfer of ownership.

% Company $c$ approaches university $u$ to verify credentials of student $s$. Since the university is the primary owner of the transcripts and degree certificates, we can assume that it can allow verifications of credentials without a separate consent from the student-- unless there is a data protection regulation that requires the contrary. In some regulations the consent of the Data Subject (DS) is necessary when the Data Owner (DO) is different from the Data Subject. Here too, data sharing only providing means access privilege without any transfer of ownership. 
\item[Job offer:] As part of the job offer, Company $c$ may require $s$ to \textit{pledge} her academic certificates as collateral, restricting their use for other job applications. Like a pledged asset, $s$ retains access but cannot transfer or re-pledge them. Here, the act of sharing takes on new semantics-- that of \textit{pledging ownership}.

% When company $c$ makes a job offer, it requires student $s$ to \textit{pledge}, or keep as \textit{collateral}, her credentials with the company so that they may not be used for applying for another full-time job as long as the current job is still active. Pledging an asset would still grant access to the owner of the asset (for instance, a car owner who pledges the car for a loan, can still continue to use the car and possess its ownership documents). However, the owner may not transfer or sell a pledged asset, and may not pledge it again elsewhere. Here, the act of sharing takes on new semantics-- that of pledging ownership.
\item[Job Contract:] Upon pledging the academic credentials by $s$ to company $c$, company $c$ becomes the \textit{pledged owner} of the credentials with limited rights, while university $u$ remains the \textit{primary owner}. Student $s$ continues to be the \textit{conferred owner} of her transcripts, but in a constrained manner, since the transcripts are pledged.

% When student $s$ gives consent to pledge his/ her credentials with company $c$, the act of \textit{pledging}, necessitates a \textit{binding agreement}, where the company and the student agree to restrictions of performing further transactions with the pledged credentials, until the duration of the job contract. During the pledge period, the student only has a copy of the credentials in his/ her custody. Despite being a \textit{conferred owner} of his/ her transcripts, after pledging, company $c$ is the \textit{current owner}, with some exclusive rights over the documents, if the student violates the terms of the contract. The pledged credentials, which is in custody of the company inherits conditions of use from University $u$, as the granting authority or the \textit{primary owner} is still the university.
\item[Contract Termination:] Once the job contract ends or the student $s$ and company $c$ agree to mutually terminate the contract, the transcripts are returned to the student, and company $c$ no longer has access to the pledged documents. The student $s$ is also longer subject to the restrictions of the pledge. The pledge is now said to be \textit{reverted}.
\end{description} 

The above represent several nuances in which ownership and consent interoperate among different independent players in a DPI. In the rest of this section, we will propose relevant computational structures that can support the use cases mentioned above. 

\subsection{Connection between Lockers}

\begin{figure*}[h]
    \centering
    % First image
    \includegraphics[width=0.65\textwidth]{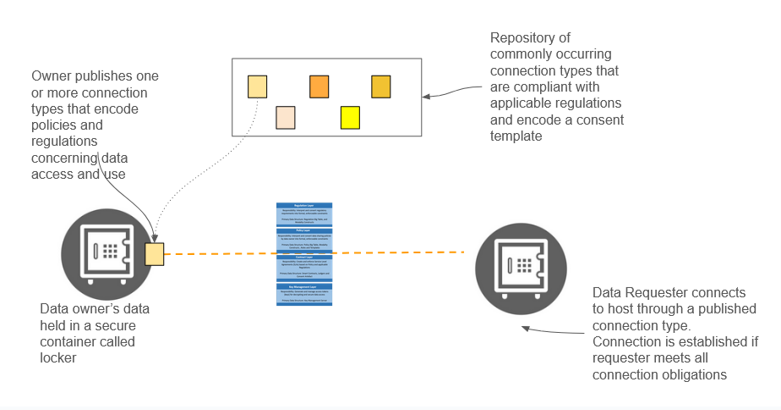} % Replace with your image
    \hspace{0.01\textwidth} % Horizontal space between the images
    % Second image
    \includegraphics[width=0.3\textwidth]{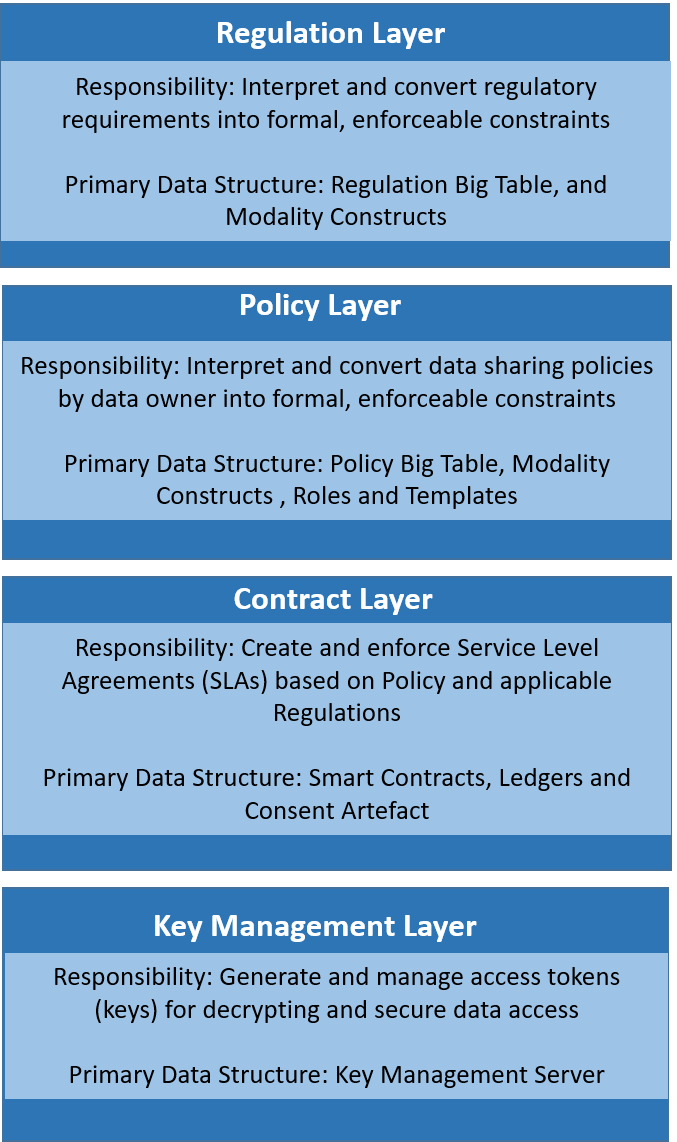} % Replace with your image
    \caption{Establishing a consensual pathway governed by four layer architecture of consent~\cite{balambiga2024cods}}
    \label{fig:sidebyside}
\end{figure*}

Before performing any data transaction, agents need to establish a \textit{connection}, between their lockers, which formalizes key properties of the transaction. The connection may be closed once the relevant artifacts are exchanged. To enable a connection, a DO uses one of its lockers to publish one or more \textit{connection endpoints} that represent a set of connection terms. A DR visiting a locker of a DO can see the list of published connection endpoints, and choose to connect any of her lockers to a given endpoint. Figure~\ref{fig:sidebyside} illustrates the establishment of a connection between the DO and DR. 

\paragraph*{Connection Type} 
A \textit{connection type} is a schema that specifies terms and conditions, under which a connection can be established between the lockers of two agents. A connection type encodes rules derived from the data access policy of the DO, as well as applicable data regulation policies. A connection type can be constructed by using one or more \textit{templates} published in a template library that encode rules governing disparate data sharing regulations. The creator of a connection type can also add more rules based on the DO's policies. Rules encoding regulation and policy constructs are called ECMA (Event-Condition-Modality-Action) rules~\cite{balambiga2024cods} that enforces three kinds of normative modalities abbreviated as $OPF$ (Obligated, Permitted, and Forbidden) constructs. An ECMA rule of the form $(e,c,m,a)$ is interpreted as: When event $e$ occurs, if condition $c$ holds then action $a$ gets a modality $m$, where $m \in \{O,P,F\}$. If event and condition are not specified in a rule, then the modality for the given action is assumed to apply at all times. Actions can include different possible actions like data sharing, purpose specifications, downloading, resharing, etc. 

Once a connection is established, all the obligations specified in the connection template need to be fulfilled to enable data exchange. A connection is said to be \textit{live} only after its obligations are fulfilled. For instance, student $s$ connecting to university $u$ to obtain her transcripts, may be required to share her college ID before the transcripts can be issued. Once this and any other specified obligation is met, data can be exchanged on this connection. After data exchange on a connection is complete, a connection can be \textit{closed}. 

%%In addition to ECMA rules that are applied prior to granting access, a connection may also encode \textit{post conditions} that need to be enforced after access is granted. In the example described earlier, there are several post conditions that need to be enforced. For instance, a data element that is \textit{conferred} on a DR, should make the DR as the new owner, but should also enforce restrictions on modification. Similarly, pledging a data element should require some restrictions to be enforced until the pledge is released. 

%To address these requirements, a new data structure called the X-node is introduced, that is detailed in the next subsection. 

% \begin{figure}
%     \centering
%     \includegraphics[width=0.52\textwidth, height=4cm]{life-cycle.png}
%     \caption{Connection Life Cycle}
%     \label{fig:lifecycle}
% \end{figure}

\begin{figure*}
    \centering
    \includegraphics[width=0.8\textwidth, height=4.2cm]{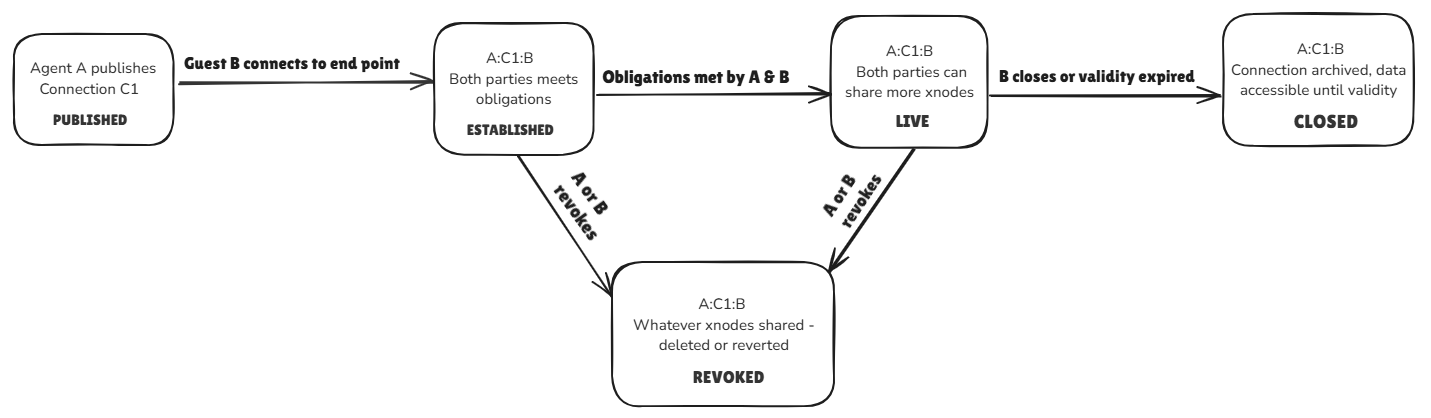}
    \caption{Connection lifecycle illustrating Agent B establishing a connection of type C1 published by Agent A}
    \label{fig:lifecycle}
\end{figure*}

\subsection{X-nodes}

% \lstdefinelanguage{json}{
%     basicstyle=\ttfamily\scriptsize, % Smaller text size
%     stringstyle=\color{red},
%     keywordstyle=\color{blue},
%     commentstyle=\color{gray},
%     numbers=none, % Disable line numbers
%     showstringspaces=false,
%     breaklines=true,
%     frame=none,
%     morekeywords={true,false,null},
%     literate=
%      *{:}{{\textcolor{blue}{:}}}{1}%
%       {,}{{\textcolor{blue}{,}}}{1}%
%       {\{}{{\textcolor{blue}{\{}}}{1}%
%       {\}}{{\textcolor{blue}{\}}}}{1}%
%       {[}{{\textcolor{blue}{[}}}{1}%
%       {]}{{\textcolor{blue}{]}}}{1}%
% }

% % Reduce spacing
% \lstset{
%     aboveskip=0.2pt, % Space above the listing
%     belowskip=0.2pt, % Space below the listing
%     xleftmargin=0.2pt, % Reduce left margin
%     xrightmargin=0.2pt % Reduce right margin
% }

Once a connection is live, data can flow on the connection. However, as noted earlier, the unit of consent is called \textit{artifact} that represents a relevant data resource. An artifact contains relevant meta-data describing the resource, details of the artifact ownership, as well as the terms of consent post sharing. To represent an artifact, we propose a data structure called X-node that in turn is modeled as a dictionary of several key-value pairs. An X-node acts both as a consent artifact that encapsulates details of the consent granted, as well as a smart contract, that enforces post-conditions of the consent after access has been granted. 

The term X-node is a generic term that represents any one of three basic types of nodes called the \textit{i-node} (``information'' node), \textit{v-node} (``virtual'' node) and \textit{s-node} (``shadow'' node). The i-node (similar to the i-node of Unix systems) represents the primary location of a data resource. The primary owner of an i-node has full authority over the data resource, including the ability to control access, modify, share, or delete the data at their discretion. The i-node also contains relevant methods to extract the resource from the resource service, and provide it to a Data Requester (DR), once a DR has legitimately gained access to the i-node. 

When a data resource is conferred on another agent, the recipient receives an \textit{s-node} also called a ``shadow'' node, for the resource. An s-node can only access the resource for reading, but cannot modify or delete the resource. The s-node contains a pointer to the original artifact (i-node or s-node) of which it is a shadow. A shadow can be created from a primary (i-node) or a conferred (s-node) resource. The original artifact also contains a list of all s-nodes that are pointing to it. 

Similarly, when an access privilege is granted for a data resource, the DR receives a ``virtual'' node or a \textit{v-node}. A v-node can be created by an i-node or an s-node, and as well as by another v-node. The v-node does not have a local copy of the resource, and only has a pointer to the original artifact. All data access through a v-node should go through the original artifact, and a v-node cannot access the resource service directly. Every v-node also has a mandatory validity field that defines the expiry date and time of the access privilege. 

%The x-nodes are called ``artifacts'' that represent logical units on which consent related policies are enforced. The actual data elements they represent are called ``resources''.
A resource (like a document), may have multiple artifacts associated with it, each pointing to different subsets of the resource. This is enabled by an abstract methods called \texttt{subset} associated with an x-node, that can be sub-classed for implementing different ways of carving subsets from resources. For instance, if consent is given for accessing only specific fields in a dataset, a separate i-node is created that represents only these fields in the resource, which is then subject to consent-related policies. If such an artifact created from a \texttt{subset} operation were to be transferred by the DO to a DR, then the DR becomes the owner of the \textit{artifact} representing a subset of the original resource. The DR will not able to access the rest of the resource, while the DO will still have access to the entire resource by virtue of owning the original artifact that represents the entire resource. The DR may create further subsets from the i-node that it owns depending on whether \texttt{subset} is enabled as a post-condition. With the x-node descriptions, we next explain different kinds of data sharing modes and consent related operations enabled by x-nodes.
%This can enable a slightly higher precedence that the creator of an artifact has over its subsequent primary owners.

    \paragraph{\textbf{Representation of X-NODES}} Each X-node is characterized by a set of essential fields as depicted in Table~\ref{tab:x-nodes}. The \texttt{creator} field identifies the agent responsible for the instantiation of the X-node. In typical scenarios, the Data Owner (DO) serves as the creator for both i-nodes and v-nodes, unless ownership is subsequently transferred. In contrast, for shadow nodes (s-nodes) generated in collateral arrangements, the Data Requester (DR) acts as the creator. The creator maintains full authority over the consent and its associated post-conditions, which are immutable by other parties; further discussion of consent transfer and collateral is provided in Section~\ref{sec:Share}, where the \textsc{TRANSFER} and \textsc{COLLATERAL} operators are defined. The fields \texttt{primary\_owner} and \texttt{current\_owner} are critical for establishing and updating possession of access rights; the specific mechanisms by which these fields are modified are addressed in the forthcoming definitions of the four core data sharing operators: \textsc{SHARE}, \textsc{CONFER}, \textsc{COLLATERAL}, and \textsc{TRANSFER} (see Section~\ref{sec:Share}). The \texttt{shadows\_list} field, present in both i-nodes and s-nodes, retains references to all related s-nodes, enabling the tracking of conferred and collateral (shadow) ownership and facilitating processes such as revocation and restoration of access. The \texttt{v-node\_list} field records all associated v-nodes, each representing a distinct access privilege. For i-nodes and s-nodes, the \texttt{pointer\_to\_resource} field indicates the location or address of the underlying data resource. The \texttt{pointer\_to\_original} field, found in v-nodes and s-nodes, links the node to its originating i-node, thereby supporting provenance and comprehensive audit trails. The \texttt{validity} field determines the temporal scope of the access rights provided by the node. The \texttt{purpose} field articulates the intended context or rationale for the node’s creation and data sharing activities, supporting principles of purpose limitation and informed consent. Finally, the \texttt{provenance} field documents the node’s history and custodial changes, thus ensuring traceability and fostering trust. 
    
\paragraph{Explanation of Essential Post-Conditions:}
The essential post- conditions of X-nodes specify the actions that the Data Requester can do with the X-node if the specific condition is checked true. The \texttt{transfer} post-condition enables \texttt{current\_owner} to transfer X-node to another agent provided the x-node is unlocked; the precise semantics of the \textsc{TRANSFER} operator are detailed in Section~\ref{sec:Share}.  The \texttt{confer} post condition is applicable for unlocked i-node or s-node (when checked true) provides \texttt{current\_owner} ability to confer to another agent. \texttt{Share} when checked true in post-conditions enables further sharing of resource access rights, allowing multiple parties to access the resource through a context of connection. The DO has access to these v-node list and can revoke access if warranted. The \texttt{collateral} post-condition allows the X-node to be pledged supporting use-cases involving temporary rights transfers. The \texttt{subset} action permits the exposure of only a restricted and relevant portion of the resource, aligning with principles of data minimization and selective disclosure. Lastly, the \texttt{download} post-condition grants the ability to obtain a copy of the resource, if permitted, under appropriate governance and audit measures. These post-conditions provides Data Owner granular control over subsequent data usage leading to a flexible yet robust framework for managing digital resource usage. The essential fields and post conditions of X-node are highlighted in Table~\ref{tab:x-nodes}. The operational definitions of the post-conditions (\textsc{SHARE}, \textsc{CONFER}, \textsc{COLLATERAL}, and \textsc{TRANSFER}) are provided in Section~\ref{sec:Share}.

% \begin{table*}

%   \caption{X-node Table}
%   \label{tab:x_node_table}
%   \begin{tabular}{|p{0.5in}|p{2in}|c|c|c|}
%   \hline
%     Node\_Id & \centering Node\_metadata &  APD & Type & Validity\\\hline
    
%     \centering 10 & \begin{lstlisting}[language=json]
% {
%     "resource_location": "file:2025/transcripts.pdf",
%     "primary_owner":"University"
%     "current_owner":null
%     "terms":<post_conditions>
% }
% \end{lstlisting} & University: Transcripts & INODE & 2025-10-19 06:42:34
%  \\\hline
%     \centering 25 & \begin{lstlisting}[language=json]
% {
%     "node_pointer": "s_node_id_54",
%     "terms":<post_conditions>
% }
% \end{lstlisting} & Company: Job Portal & VNODE & 2025-10-19 06:42:34\\\hline
%     \centering 54 & \begin{lstlisting}[language=json]
% {
%     "node_pointer": "i_node_id_10", 
%     "resource_pointer": "file:2025/transcripts.pdf",
%     "primary_owner": "university",
%     "current_owner": "student"
% }
% \end{lstlisting} & Student: Education Docs & SNODE & 2025-10-19 06:42:34\\\hline

%   \end{tabular}
% \end{table*}

\begin{table*}[h]
\begin{tabular}{|c|p{1.2in}|p{2.2in}|p{1.2in}|}
    \hline
     \textbf{X-node} & \textbf{Function} & \textbf{Essential fields} & \textbf{Essential post-conditions}\\
     \hline
     \hline
     i-node & ``information''  node: represents primary location of resource & \texttt{creator}, \texttt{primary\_owner}, \texttt{current\_owner}, \texttt{shadows\_list}, \texttt{v-node\_list}, \texttt{pointer\_to\_resource}, \texttt{purpose}, 
     \texttt{provenance} & \texttt{transfer}, \texttt{confer}, \texttt{share}, \texttt{collateral}, \texttt{subset}, \texttt{download}\\
     \hline 
    v-node & ``virtual'' node: represents an access privilege to an i-node & \texttt{creator}, \texttt{current\_owner}, \texttt{pointer\_to\_original}, \texttt{validity}, \texttt{v-node\_list}, \texttt{purpose},
    \texttt{provenance} & \texttt{transfer}, \texttt{share}, \texttt{download}\\
    \hline 
    s-node & ``shadow'' node: represents a conferred or pledged ownership & \texttt{creator}, \texttt{primary\_owner}, \texttt{current\_owner}, \texttt{pointer\_to\_original}, \texttt{shadows\_list}, \texttt{v-node\_list}, \texttt{pointer\_to\_resource},\texttt{purpose}, \texttt{provenance} & \texttt{transfer}, \texttt{share}, \texttt{collateral}, \texttt{subset}, \texttt{download}\\
    \hline
\end{tabular}
\caption{Essential fields of X-nodes}
\label{tab:x-nodes}
\end{table*}

\section{Data Exchange and Consent Operations}\label{sec:Share}

\begin{figure*}
    \centering
    \begin{tabular}{cc}
        % Top row
        \begin{minipage}{0.45\textwidth}
            \centering
            \includegraphics[width=0.6\textwidth]{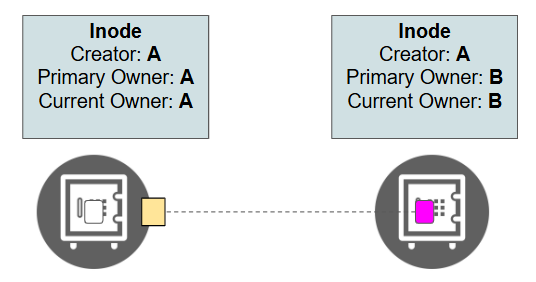}
            \\[0.3em]
            \textbf{(i)}
            \\[1em]
        \end{minipage}
        &
        \begin{minipage}{0.45\textwidth}
            \centering
            \includegraphics[width=0.6\textwidth]{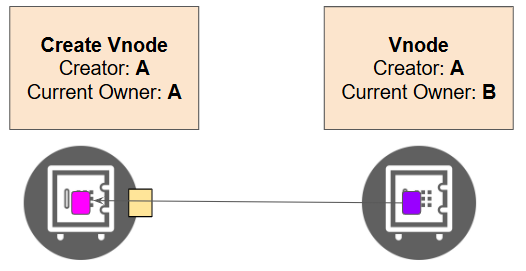}
            \\[0.3em]
            \textbf{(ii)}
            \\[1em]
        \end{minipage}
        \\[5em]
        % Bottom row
        \begin{minipage}{0.45\textwidth}
            \centering
            \includegraphics[width=0.8\textwidth]{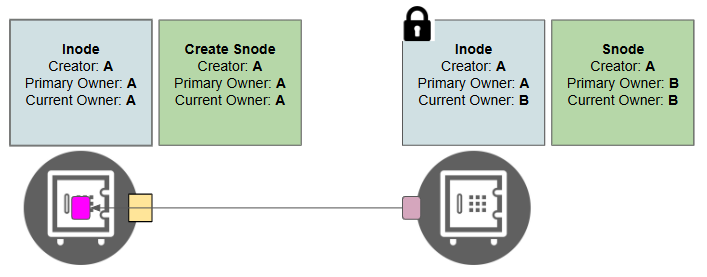}
            \\[0.5em]
            \textbf{(iii)}
        \end{minipage}
        &
        \begin{minipage}{0.45\textwidth}
            \centering
            \includegraphics[width=0.8\textwidth]{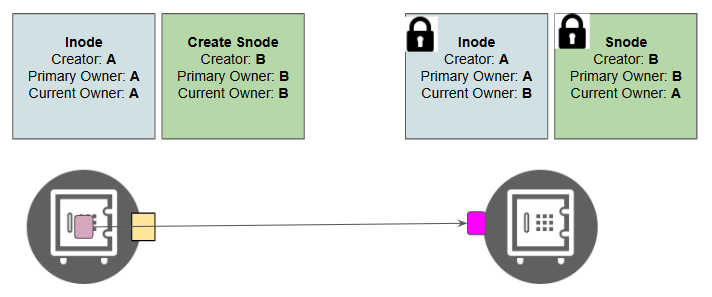}
            \\[0.5em]
            \textbf{(iv)}
        \end{minipage}
    \end{tabular}
    \caption{Sharing Operators for Transition of Ownership (i) TRANSFER (ii) SHARE (iii) CONFER (iv) COLLATERAL}
    \label{fig:sharing_operators}
\end{figure*}
As noted in the motivating example, data sharing in a DPI can take on various interpretations. In this section, we will introduce different modes of data exchange protocols and show how they can be supported using the x-nodes introduced above.

\paragraph*{\textbf{SHARE:}}

The first kind of data exchange operation we introduce is the SHARE operation. In this mode of sharing, the DO shares an \textit{access privilege} to a resource $r$. The DO has an i-node (identified as $do.i(r)$) corresponding to the resource to be shared, that lists the DO as the primary owner. A v-node is then created by the DO that points to the i-node of the resource, and its owner set to the Data Owner (DO). A validity period is also set for the v-node. When the v-node is transferred to the DR and accepted by the DR, the owner is set to DR. Let this v-node be identified as $dr.v(do.i(r))$.  

As long as the v-node is valid, the Data Recipient (DR) has access privilege to the i-node within the Data Owner’s (DO) locker, thereby obtaining access to the underlying resource. Only an i-node or s-node can access the resource server that provides access to the resource. The resource server that gets an access request also receives the \textit{access tunnel} called $dr.v(do.i(r))$ through which the resource is sought. In an access tunnel of the above form, the first element ($dr.v$ in this case) is called the \textit{origin} of the request, and the last element ($do.i$) is called the \textit{ground} of the request. The resource server will approve the access only if $do$ is the \texttt{primary\_owner} of $do.i(r)$. Also, if the origin of the request does not correspond to the ground of the request, then the resource server does not allow any other operations other than \texttt{read} on the resource, like \texttt{edit} or \texttt{delete}. 

Since every access via a share comes through the i-node of the owner and is logged, the DO is aware of every data access made by the DR for the resource. Figure~\ref{fig:sharing_operators}(ii) schematically depicts a SHARE operation with the i-node shown in pink and the v-node in purple.

\paragraph{\textbf{CONFER:}} 

The second kind of data data exchange operation is \textit{conferment} operation. This kind of operation is suitable for cases where the DO shares an immutable copy of the data to the DR, for which, the DR becomes the owner. Examples include issuing of certificates, licenses, permits, etc. Here, the primary owner is the issuing authority, and the conferred owner is the recipient. 

To enable this, the DO, which has an i-node for the resource, creates an s-node which is then sent to the DR. The \texttt{primary\_owner} (PO) and \texttt{current\_owner} (CO) fields of the s-node are set to the DR, which enables the DR to function as the full owner of the s-node. The \texttt{primary\_owner} of the i-node continues to be the DO, while the \texttt{current\_owner} is set to DR. Figure~\ref{fig:sharing_operators}(iii) schematically depicts a CONFER operation with the i-node shown in pink and the s-node shown in brown. When the PO and CO fields of a x-node (i-node or s-node) are not the same, then the x-node is said to be \textit{locked}. A locked x-node cannot be transferred, conferred or pledged to others. Hence, when a resource (like a certificate or license) is conferred on someone, the same resource cannot be conferred to someone else as indicated by the locked icon on the i-node in the figure. Only \texttt{primary\_owner} of a resource possessing an unlocked i-node can confer the resource. Although a conferred owner can directly access the resource through the s-node, it is a read-only access. Any edits on the resource should be performed by the \texttt{primary\_owner} of the resource, through the i-node. The i-node also maintains a list of s-nodes pointing to it, which are then notified whenever a change is made. Since the s-nodes can directly access the resource server, their subsequent access will now access the new version of the resource. Unlike the case of a v-node, the owner of an s-node can can directly access the resource without going through the i-node. This means that the primary owner is not notified of each access by the conferred owner. 

A conferred ownership can also be seen as a form of ownership delegation with limited privileges. For instance, a Power of Attorney (PoA) can be implemented using a conferment, where relevant documentation can be provided to a third party who can then present it to different agencies to avail services and perform transactions, without the ability to make any changes to the documentation. 

\paragraph{\textbf{TRANSFER:}} 

This mode of data data exchange operation involves a transfer of ownership from the DO to the DR. In this case, the x-node of the resource itself is transferred to the DR and both its PO and CO fields are set to the DR. Transfer can be performed on i-nodes, v-nodes and s-nodes. A transfer of ownership provides the DR with complete and unrestricted ownership of the resource, and the former DO will not have any access to the x-node after transfer. Figure~\ref{fig:sharing_operators}(i) depicts a TRANSFER from the DO to the DR. The \textit{i-node} of the resource (depicted in pink) is removed from the DO's locker and is transferred to the DR's locker, with the new owner information updated in the i-node. This operation guarantees policy independence for the new owner, with the former owner having no say in how the artifact is used after the transfer. The new owner can now also define new usage policies independently. 

Transfer of an x-node also invalidates v-nodes and s-nodes pointing to it, as a transfer may also involve changes in policy under the new owner. Any existing v-nodes and s-nodes will have to be regenerated and re-issued under the policies of the new owner. The invalidation is performed by maintaining a list of incoming links from s-nodes and v-nodes. Transfer also updates the \textit{provenance} field in the x-node upon transfer. This enables tracing the path of a given artifact in the DPI through transfer operations. 

While TRANSFER involves relinquishing ownership over the artifact by the DO, if the DO is the creator of the artifact (represented by the \texttt{creator} field), there are still some post-conditions that can prevail over transfer operations. Any forbidden post-condition set by the creator of the artifact, continues to be forbidden across transfers. 
%This constraint is further clarified later in this section when we differentiate between a resource and an artifact. 

\paragraph{\textbf{COLLATERAL:}}

In this mode of data exchange operation, the data resource is pledged as a collateral by the DO to the DR. Examples include an applicant providing her car documents as collateral to a bank to avail a loan, or a job applicant pledging her university degree and transcripts in response to a full-time job offer. Pledging comprises of the following steps: (i) The DO first sends the i-node of the pledged resource, to the DR, keeping the \texttt{primary\_owner} (PO) as DO itself, and setting the \texttt{current\_owner} (CO) to DR (ii) Upon receiving the i-node, the DR creates an s-node for this, with the \texttt{primary\_owner} as DR, and \texttt{current\_owner} as DO.  

We can note that in the above operation, both the i-node and s-node are locked as a result of pledging as shown in Figure~\ref{fig:sharing_operators}(iv). The \texttt{primary\_owner} of the i-node is DO, while the DR is only keeping the i-node in confidence as a pledged owner. In return, the DR issues an immutable shadow to the DO, for which, DR is the \texttt{primary\_owner}, and DO is entitled to use it as the \texttt{current\_owner}. Since both x-nodes are locked, neither the DR nor DO can further pledge, confer or transfer these artifacts until the pledge is released. Pledging can also be performed on s-nodes that are unlocked. In the example of an applicant applying for a job, the degree credentials of the applicant is an s-node. Unlike pledging, conferment does not lock the s-node and both PO and CO are set to the degree holder. As a result, this can be pledged with the same semantics as that of an i-node.

\paragraph{\textbf{Compositions of Data Exchange Operations:}}

The four types of Data Exchange Operations described above, can be composed in different ways. 

A v-node that is created due to a SHARE operation, can be further subject to a SHARE (provided it is permitted by the post-conditions). This creates yet another v-node pointing to the original v-node. This can be used to create \textit{access tunnels} where a resource can be accessed through a legitimate, consensual pathway comprising of several layers of consent. A data access request can originate from any of the interim nodes in such a tunnel, but the final request to the resource server goes from the i-node (or s-node) residing at the DO's locker. The DO can also get an idea of how this shared resource has been shared further, and the different entities that are accessing it. A v-node can be prevented from further sharing by setting a post-condition to this effect, at the time of creation. 

\begin{figure*}
   \centering
   \includegraphics[width=0.7\textwidth]{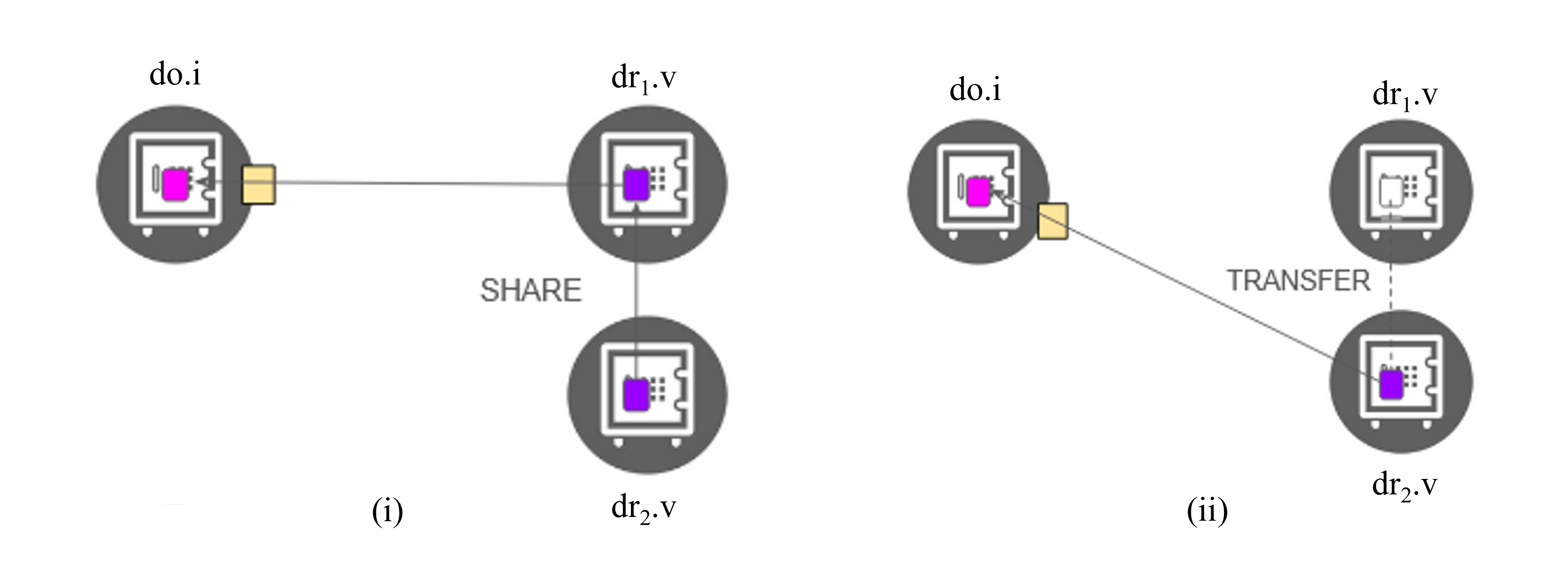}
   \caption{(i) An \textit{access tunnel} established for origin of request ($dr_2$) to the ground of the request ($do$) after performing SHARE of a SHARE (ii) A v-node ($dr_1.v)$ subject to TRANSFER from $dr_1$ to $dr_2$}
   \label{fig:access_tunnel}
\end{figure*}

A v-node can be subject to a TRANSFER provided that it is permitted by the post-conditions. When a v-node is transferred by a DR to another agent (say DR2), then the earlier DR loses the access pathway to the resource, which is now held by DR2. The share vs transfer of v-node is shown in Figure~\ref{fig:access_tunnel}. The operations CONFER and COLLATERAL are not defined for v-nodes.

A s-node that is the result of a conferment can be further subject to SHARE, COLLATERAL or TRANSFER. In some cases, the primary owner may disallow TRANSFER as a post-condition at the time of conferment (as in the case of conferred degrees) and hence cannot be further conferred. Locked artifacts that are a result of COLLATERAL operations, cannot be further transferred, conferred or pledged. However, SHARE operation is still defined on them (unless prohibited by post-conditions), where access tunnels can be established to locked x-nodes through v-nodes. 

\paragraph{\textbf{CLOSE, REVOKE and REVERT:}}

In the proposed model, consensual data exchange consists of two levels of consent-- consent for establishing a connection, and the consensual sharing of specific artifacts. This gives rise to multiple ways in which consent is deemed closed or revoked. 

The CLOSE operator defined on connections results in closing an open connection. This is typically invoked after the successful completion of data sharing. Artifacts downloaded by the DR from this connection continue to remain with the DR. Some artifacts like v-nodes have a validity period, after which they can no longer be used. However, until the validity period, the DR can continue to access the primary i-node (and the resource), even after the connection is closed. 

The REVOKE operator is also defined on a connection, and results in a \textit{rollback} of the connection. %All v-nodes exchanged through this connection are deleted.
All i-nodes and s-nodes created through this will be reverted to the original owner by resetting the primary and current owners to their earlier values. The REVOKE operator is also defined over artifacts that are shared or transferred. Revoking a SHARE entails deletion of the v-node that was created as part of the SHARE. 
%Revocation has a cascading effect. Any further SHAREs of the v-node are also deleted. 
Revoking a TRANSFER entails bringing the transferred x-node to its original location and resetting its owner fields. Revocation of artifacts can only be performed on a connection that is still open and live. Revoking an artifact does not close or revoke the connection. 

For artifacts that are exchanged through a CONFER or COLLATERAL operator, another operator called REVERT is defined that can be used even after the connection is closed. Conferment and pledging are long-term exchanges, and the connection through which they are exchanged may be closed after the exchange. The REVERT operator reverses a pledge by bringing back the pledged i-node to its original owner and setting both original and current owner to the DO. Similarly, a REVERT operator on a conferred artifact results in annulment of the conferment and deletion of the corresponding s-node. While the primary owner can unilaterally revert a conferment, revoking a collateral needs to be approved by both parties.

\section{Adversarial Scenarios}

In this section, we evaluate the efficacy of the consent model in the DPI architecture, through adversarial scenarios that attempt to undermine the integrity of the system, and detail how the proposed system responds to these.

\paragraph*{Scenario 1.} \textit{Consider a user, Alice, pledging an artifact using COLLATERAL to Bob for temporary custody. Bob attempts to exploit his ownership privilege, by trying to update the contents in the resource or sell the artifact by performing a TRANSFER of the i-node.}

\textbf{Scenario Handling:} When Alice performs a COLLATERAL to pledge her resource artifact, Alice remains the primary owner on the artifact, and Bob is the current owner. This sets the state of the i-node to \textit{locked}. A locked i-node cannot be transferred or used to modify the resource. Hence, Bob will not be able to exploit any ownership privilege on the resource even with the possession of the pledged i-node. However, Bob will be able to read the contents of the resource, and also create a v-node pointing to the pledged i-node and share the same to be read by others. The latter can be prevented by Alice setting the \texttt{share} post-condition to false, before pledging the artifact. 
%she retains the \textit{s-node}, which is an immutable copy of the data artifact. This can essentially act as the reference point for verification and can be audited against, if the resource artifact is altered by Bob. To ensure responsible custody of the \textit{i-node} by Bob, restrictions of performing ownership transitions like TRANSFER or CONFER can be specified as part of the active connection.\newline

\paragraph*{Scenario 2.} \textit{An adversary creates a locker mimicking that of a legitimate agent, to gain access to some sensitive artifacts shared by another agent. For example, consider Alice publishing a connection endpoint for her company, inviting potential vendors for a given tender call. Upon connecting, the vendors receive sensitive details of the project for which they are bidding. Bob creates a locker representing a fictional company, and connects to this endpoint, in order to access sensitive project information.}

\textbf{Scenario Handling} Such a scenario should be handled by judicious creation of the connection terms. Vendors connecting to the company connection endpoint should be obligated to prove their credentials by sharing relevant statutory documentation. The connection can be made live only after verification of the obligated documentation shared by Bob. The semantics of connection obligations is that, the connection is not live and no data can be exchanged on the connection, until all the obligations are met. 

%If the credentials of Alice are compromised, Bob can make a consent request to the hospital for accessing personal health data of Alice. To safeguard an illegitimate exchange, the hospital can, as part of it's \textit{connection-type} obligate the Data Requester to submit identity proofs with latest attestations before granting the consent request. These safeguards can be encoded in the connection as per the compliance requirement of the hospital. Secondly, if Alice is aware of her credentials being leaked, she can immediately inform a moderator to block all actions of the APD or setup of new connections.\newline

\paragraph*{Scenario 3.} \textit{An adversary with access to a v-node, attempts to replicate the resource outside the DPI, bypassing its access pathways. For instance, a third-party entity leaks data after clicking photographs from another device after the entity gets a v-node from the Data Owner}.

\textbf{Scenario Handling:} In this scenario, the Data Requester (DR) has violated the constraints of immutability and purpose of use obligation that has been agreed with the Data Owner as part of his connection, through which the v-node was shared. The client application or the browser through which the resource artifact is viewed, can enforce the constraints of \textit{view-only}, by disabling screenshots and download of the resource artifact. However, the consent manager cannot directly control or monitor external physical actions like taking photographs. The resource can be overlaid by watermarking, that makes any unauthorized photographs traceable. The consent manager would also have logged the legal capacity in which the Data Requester made the access, if the entity is found to be faulty after later external investigations.

\paragraph*{Scenario 4.} \textit{A Data Requester, which is in custody of a v-node, shared by the Data Owner, tries to cascade SHARE to other third party entities}

\textbf{Scenario Handling} In this case, if the original intention of the Data Owner is to not give consent to further re-sharing, then the \texttt{share} post-condition on the v-node should set to false by the DO, before transferring it to the DR. Even when a v-node is subject to further shares, it only creates an access tunnel whereby, the resource can only be accessed through the i-node that is in possession of the DO. The DO can revoke consent by disabling the access pathway of the respective cascaded v-node and any additional access pathways at any time. 

\paragraph*{Scenario 5.} \textit{A Data Requester (DR) is located in a legal jurisdiction different from the regulatory requirements of the connection published by a Data Owner (DO).}

\textbf{Scenario Handling} This constitutes a \textit{cross-border} data exchange, which are not supported by the framework in its current form. Cross-border data transfers are facilitated by regulations governing both the sending and receiving of data across jurisdictional boundaries. In the current model, both DR and DO are assumed to be within the same jurisdictional boundary, whose regulations are encoded in the connection terms. 

\section{Discussion}

A key requirement for any robust Digital Public Infrastructure (DPI) is the ability to address the security and governance needs as data transitions through its core states: \emph{at rest}, \emph{in transit}, and \emph{in use}. Our architecture, through the use of regulated lockers and auditable connections, directly models and secures these states: lockers protect data at rest via strong access controls, while connections govern and trace all data in motion and use, ensuring accountability and adherence to intended purposes throughout the data lifecycle.

The introduction of connections as regulated, obligation-bound channels not only enforces lawful and contextual data exchange but also empowers Data Owners with fine-grained ongoing control. For example, Data Owners can dynamically limit the scope (e.g., read-only access, validity window), monitor usage through real-time v-node tracking, and revoke privileges in response to misuse—providing a tangible means to exercise rights mandated by privacy legislation. In contrast to approaches that are confined to single-website consent banners, our method enables network-wide, scalable consent management and traceability within Digital Public Infrastructures (DPIs). This design not only meets but surpasses the baseline requirements of regulations such as GDPR and other data regulations, offering comprehensive auditability and enhanced user agency across diverse platforms and services.

While this paper primarily focuses on the architectural formalization, a practical implementation can be realized within platforms such as DigiLocker~\cite{digilocker_gov_in}, which serves as a secure, government-backed digital document wallet enabling citizens to store and share authentic electronic records. In such a deployment, data ownership and all associated operations may be managed by a dedicated Consent Service utilizing X-nodes, which encapsulate real-world data sharing semantics and provide comprehensive end-to-end traceability of both consent and ownership transitions throughout the data flow lifecycle. Consent artifacts, as conceptualized, can be administered and exchanged using standardized, machine-readable formats such as JSON facilitating interoperability across digital systems. Enforcement and management of these artifacts leverage privacy-by-design mechanisms established in initiatives like India Stack~\cite{indiastack_identity} and MyData~\cite{hyysalo2016consent}. Prospective work will focus on deploying the proposed model within an operational Digital Public Infrastructure (DPI) gateway, systematically examining end-user understanding and agency, and conducting adversarial and other system-level tests to rigorously validate usability and resilience.

However, several challenges and opportunities for further research remain. The effectiveness of user-facing consent dashboards and real-time revocation tools must be validated through user studies to understand if this truly improve end-user agency and understanding. Scalable enforcement of fine-grained obligations in high-traffic and adversarial scenarios also warrants rigorous evaluation. In summary, by abstracting consent through X-node and integrating continuous enforcement into the data flow, our design addresses critical gaps in current consent management. This empowers individuals, supports compliance, and lays practical groundwork for secure, user-centric data exchange at DPI scale.

\section{Conclusions}

%A data owner can append additional granular data sharing policies along with the regulations by the state that are encoded in the system. 
A comprehensive design of Digital Public Infrastructures is necessary to address conflicting requirements of upholding individual and organizational autonomy, as well as upholding public interest and security. Given the evolving landscape of data protection laws, it is becoming infeasible for individuals and even organizations to manage their public data flows.  
%involves huge effort and resources by the organizations managing data and consent. 
%A DPI model proposed allows management of data exchange through organisation, government, individuals and compliance with various regulations templates ensures compliance with relevant sections of the regulation based on the type of data shared.  

In this work, we presented an architecture using X-nodes and Connections to support data exchange that empowers data owners and supports compliance with diverse regulations, including personal and other data laws. One of the key challenges in this space is that different laws and domains often require different consent processes and data management structures. Digital Public Infrastructure (DPI) provides a common and flexible foundation to address this challenge, allowing consent mechanisms and data flows to be adapted according to each legal framework's needs. By managing data sharing through regulated pipelines called connections, and by supporting detailed tracking of consent and ownership changes, our design allows data owners to retain control even after data is shared. The use of X-nodes also models different forms of ownership, such as delegated or pledged rights, making it possible to reflect the different ways consent and data control work in the real world. Overall, using DPI ensures that consent management can be both robust and adaptable, giving individuals genuine control and helping organizations remain compliant as data laws and expectations evolve.

\bibliographystyle{splncs04}
\bibliography{ref}
\renewcommand{\UrlFont}{\small\ttfamily}
\end{document}